\documentclass[prb,twocolumn,showpacs,amsmath,amssymb]{revtex4-1}

\usepackage{graphicx}% Include figure files
\usepackage{amssymb,amsmath}
\usepackage{dcolumn}% Align table columns on decimal point
\usepackage{bm}% bold math
\usepackage{color}

\begin{document}

\title{Bloch Oscillations of an Exciton-polariton Bose-Einstein Condensate}
%\date{\today}

\author{H. Flayac}
\affiliation{Clermont Universit\'{e}, Universit\'{e} Blaise Pascal, LASMEA, BP10448, 63000 Clermont-Ferrand, France}
\affiliation{CNRS, UMR6602, LASMEA, 63177 Aubi\`{e}re, France}

\author{D.D. Solnyshkov}
\affiliation{Clermont Universit\'{e}, Universit\'{e} Blaise Pascal, LASMEA, BP10448, 63000 Clermont-Ferrand, France}
\affiliation{CNRS, UMR6602, LASMEA, 63177 Aubi\`{e}re, France}

\author{G. Malpuech}
\affiliation{Clermont Universit\'{e}, Universit\'{e} Blaise Pascal, LASMEA, BP10448, 63000 Clermont-Ferrand, France}
\affiliation{CNRS, UMR6602, LASMEA, 63177 Aubi\`{e}re, France}

\begin{abstract}
We study theoretically Bloch oscillations of half-matter, half-light quasi-particles: exciton-polaritons. We propose an original structure for the observation of this phenomenon despite the constraints imposed by the relatively short lifetime of the particles and the limitations on the engineered periodic potential. First, we focus on the linear regime in a perfect lattice where regular oscillations are obtained. Second, we take into account a realistic structural disorder known to localize non-interacting particles, which is quite dramatic for propagation-related phenomena.  In the non-linear condensed regime the renormalization of the energy provided by interactions between particles allows us to screen efficiently the disorder and to recover oscillations. This effect is useful only in a precise range of parameters outside of which the system becomes dynamically unstable. For a large chemical potential of the order of the potential's amplitude, a strong Landau-Zener tunneling tends to completely delocalize particles.

\end{abstract}

\maketitle

\section{Introduction}

Cavity exciton-polaritons (polaritons) are mixed exciton-photon quasi-particles. A fundamental difference with respect to simple photons is that polaritons are interacting both with themselves and with other excitations of the surrounding media. As a result, they can exhibit Bose-Einstein condensation \cite{Kasprzak} and other related fascinating effects, such as superfluidity \cite{Amo}, vortex formation \cite{Rubo,LagoudakisV}, and Josephson oscillations\cite{Sarchi,LagoudakisJO}. A strong specificity of polariton condensates with respect to atomic ones is the possibility of performing resonant excitation experiments that allow us to achieve a very high degree of control on phenomena such as parametric instabilities \cite{Savvidis,Skolnick} or bistability \cite{Baas}. A second specificity lies in the specific spin dynamics and spin structure of polaritons \cite{Review} which has allowed the demonstration of interesting phenomena, such as the formation of half vortices \cite{LagoudakisHV} or the demonstration of multistability \cite{Gippius}. However, polaritons typically suffer both from a short life time (in the $ps$ range) and from localization induced by the presence of structural disorder, which tends to limit the observability of propagation effects. Recent progress in growth and technology enables the fabrication of very high quality samples in which polaritons can live up to $50$ $ps$. Such large values have allowed the generation of extended condensates with coherence length larger than the size of the sample \cite{Wertz}. These characteristics are opening a new research field where it will be possible to study  propagation phenomena of polariton condensates in detail. One of these phenomena is Bloch oscillations (BOs). A periodic motion of electrons within the first Brillouin zone (FBZ) of a semiconductor biased by an electric field was predicted to occur about 80 years ago by F. Bloch \cite{Bloch}. This phenomenon has been widely theoretically discussed during the twentieth century \cite{Wannier}. However, its experimental observation requires the coherent propagation distance to be larger than the oscillation period. This constraint made the observation of electronic BOs in bulk semiconductors impossible. It took 60 years and required the fabrication of artificial crystals to finally observe this phenomenon \cite{Mendez,LeoWaschkeDekorsky}.

BOs have been described and observed in three alternative systems showing extended spatial and long temporal coherence, namely, coherent light waves propagating in photonic crystals \cite{Malpuech,Agarwal}, ultracold atoms \cite{Dahan}, and atomic Bose-Einstein condensates (BECs) in optical lattices \cite{Choi,Morsh}. In the linear regime, the electronic, photonic, and atomic systems show very similar single-particle behavior. However, they differ strongly in their specific non-linear response. In electronic systems, electron-electron interaction induces some dephasing, leading to the destruction of coherent BOs. On the other hand, photons are non-interacting particles, and of course, if they propagate in a linear media, no peculiar non-linear behavior is expected. The situation is radically different for atomic BECs in optical lattices where the transport properties are strongly affected by the non-linear effects linked with the density of condensed particles. The latter topic is still widely under discussion, as rich physics emerges out of it \cite{Cristiani,Menotti,Deissler}. Indeed, while an optical lattice constitutes a perfect periodic optical crystal that can lead to long living BOs of dilute (ultra)cold atoms \cite{Ferarri}, a BEC can experience decoherence \cite{Buchleinter} and dynamical instabilities \cite{WuNiu, Kolovsky} due to non negligible interactions. Moreover, the account of disorder leads to phenomena such as Anderson-like localization \cite{AndersonSchulte,Modugno} or the breakdown of superfluidity \cite{Astrakharchik, Leboeuf}. The non-uniformity of the potential tends to prevent wave function's spreading, while interactions drive the system toward superfluidity. This interplay has serious consequences for the phase diagram of the system \cite{Roth, Damski} and most importantly, the disorder induces extra dephasing which is revealed by a damping of the matter wave's BOs \cite{Schulte,Drenkelforth}.

Currently, several works have already been devoted to the behavior of polariton condensates in periodic lattices \cite{Cho,YamamotoNature,Yamamoto,Skolnick2010,Skryabin}, and this topic has started to attract a lot of attention. In this paper we theoretically describe BOs of polaritons in a patterned one-dimensional microwire. In both atomic and optical systems, it is possible to design a potential that allows short period oscillations (less than 1 $ps$ for instance in Ref.\onlinecite{Malpuech}). This cannot short period cannot occur in polaritonic systems, where oscillations must take place in the strongly coupled part of the lower polariton branch. In total, the width of the first Brillouin zone has to be smaller than one fourth of the Rabi splitting. This gives values of the order of $1-5$ $meV$ and implies some lower bound for the BO period that should be smaller than the polariton lifetime.

In the first part of this work, we propose a realistic structure in which polariton BOs could be realistically observed and show numerical simulations in the linear regime. In the second part of the paper, we therefore study the effect of structural disorder on BOs of a polariton condensate. Indeed, an important peculiarity of the polariton system is that it is affected by intrinsic sample disorder which can be absent in atomic systems and which perturbs much less pure photons' propagation. As a result, polaritons are often found to be localized \cite{Kasprzak,LagoudakisJO}, which evidently can make the observation of propagation effects like BOs difficult. In the non-interacting regime, we can expect that the polaritons' motion will be blurred (dephased) with increasing disorder potential. So, strictly speaking, an uncondensed thermal gas of polaritons cannot undergo BOs if the disorder is too important. Nevertheless, the formation of an interacting polariton BEC should permit us to recover an oscillatory behavior as a result of a partial screening of the disorder. The growth of interaction energy leads the polariton gas toward superfluidity \cite{Malpuech2007} at low velocities, where the influence of disorder is the strongest.

In general, the oscillations are found to be damped because of both the residual scattering by disorder when polaritons are moving at a supersonic velocity, and the occurrence of specific parametric instabilities, discussed in Ref.\onlinecite{Skryabin} and well known in the atomic BEC field, when polaritons are accelerated to the inflexion point in the first Bloch band of the dispersion. The most interesting result is that one of these effects can compensate the other: The interactions protect (at least partially) the BOs from the damping linked with disorder. The precise characterization of these non-linear scattering processes is not in the scope of the present work. However, we would like to point out a key specificity of the polariton system with respect to the atomic one: the possibility to perform resonant excitation experiments (with spatial and frequency resolution) that allow us to address specifically the various non-linear scattering processes that interacting particles accelerated in a disordered periodic potential can undergo.

\section{Structure and linear Bloch oscillations}
\begin{figure}
  \includegraphics[width=0.5\textwidth,clip]{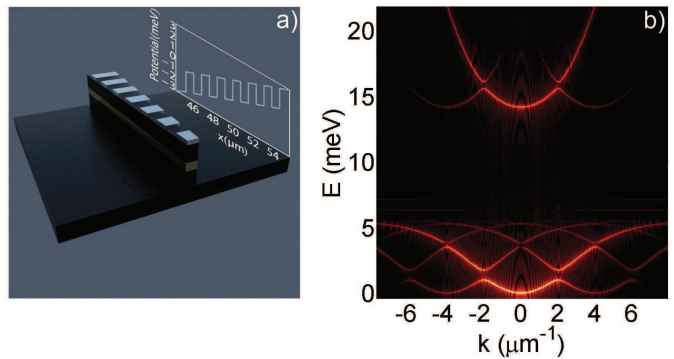}\\
  \caption{(Color online) (a) The structure: laterally narrowed wire shaped microcavity with periodic metallic depositions and the resulting total potential (Only the central part where oscillations take place is represented for clarity). (b) The associated initial dispersion of the particles modified by the periodic potential $U_{ex}(x)$ created by the metal.}
  \label{Fig1}
\end{figure}

The structure proposed is schematically shown in Fig.\ref{Fig1}(a). It is based on a $L_x=100$ $\mu m$ long GaAs microcavity etched in the $y$-direction in order to realize a wire having a lateral size in the $\mu m$ range. The confinement energy provided by the lateral etching is approximately ${E_{c}} = {\hbar ^2}{\pi ^2}/2{m^*}{L_y^2}$, where $m^* \sim 2 m_{ph}$ (at zero detuning) is the effective mass of the polariton. The linear potential ramp needed in order to mimic the electric field acting on electrons can be realized by changing the lateral size of the wire along its main axis $x$ with a square root dependency ${L_y}(x) \sim {L_0}/\sqrt x$ in the region where BOs are expected. In principle, any type of accelerating potential can be designed thanks to modern lithography technics but of course, there are some limitations due to the confinement energy dependance. It is also possible to make use of the wedge character of microcavities along the growth $z$-direction \cite{Sermage2002}. It should be noted that this potential ramp will be acting on the photonic part of the quasi-particle. The periodic potential required to open a gap in the polariton dispersion can be obtained either by depositing a metallic pattern along the wire \cite{Yamamoto}, by excitation of a surface acoustic wave \cite{Lima,Skolnick2010}, by a square-wave-like lateral etching, by a chain of coupled Josephson junctions or even by means of Tamm plasmons \cite{Kaliteevski}. The first two techniques allow the creation of a periodic potential acting on the excitonic part of the polariton, and in this work we will focus on the first one.

To describe the dynamics of the system we use a set of one-dimensional time dependent Scr\"{o}dinger and Gross-Pitaevskii equations for the photonic $\psi _{ph}(x,t)$ and excitonic fields $\psi _{ex}(x,t)$, coupled by the light-matter interaction  \cite{Shelykh2006} $\Omega_R=14$ $meV$ (Rabi splitting of realistic samples), taking into account the finite lifetimes $\tau_{ph}=40$ $ps$ and $\tau_{ex}=150$ $ps$ of photons and excitons, and a quasi-resonant Gaussian photonic (laser) pump $P\left(t\right)$:
\begin{eqnarray}
  i\hbar \frac{{\partial {\psi _{ph}}}}{{\partial t}} = &-& \frac{{{\hbar ^2}}}{{2{m_{ph}}}}\frac{{{\partial ^2}{\psi _{ph}}}}{{\partial {x^2}}} + {U_{ph}}{\psi _{ph}} \\
  \nonumber  &+& \frac{\Omega_R }{2}{\psi _{ex}} - \frac{{i\hbar }}{{{2\tau _{ph}}}}{\psi _{ph}} + P\left(t\right) \\
  i\hbar \frac{{\partial {\psi _{ex}}}}{{\partial t}} = &-& \frac{{{\hbar ^2}}}{{2{m_{ex}}}}\frac{{{\partial ^2}{\psi _{ex}}}}{{\partial {x^2}}} + {U_{ex}}{\psi _{ex}} \\
  \nonumber  &+& \frac{\Omega_R }{2}{\psi _{ph}} + \alpha {\left| {{\psi _{Ex}}} \right|^2}{\psi _{Ex}} - \frac{{i\hbar }}{{{2\tau _{ex}}}}{\psi _{ex}}
\end{eqnarray}
where $m_{ph}=5 \times 10^{-5} m_{el}$, $m_{ex}=0.4 m_{el}$ and $m_{el}$ are the photon, exciton, and free electron masses, respectively. The nonlinear term with $\alpha=6 E_{b} a_{B}^2/S$, where $E_b$ is the exciton binding energy, $a_B$ its Bohr radius, and $S$ is a normalization area, accounts for the polariton-polariton coherent repulsive interactions \cite{Ciuti1998}. Finally, $U_{ph}(x)=-Fx$ and $U_{ex}(x)$ are the accelerating ramp potential producing a constant force $F=0.2$ $meV/\mu m$ and the squarewave potential of period $d=1.56$ $\mu m$ and amplitude $A=2$ $meV$, respectively . Here we make the following remark: The barrier height is another important limitation to our structure. Indeed, contrary to optical potential where it can be easily tuned in wide ranges, for polaritons metallic depositions will tend to absorb photons, which can shorten the polariton lifetime even more. Thus, we can hardly consider large enough band gaps to allow us to employ the tight binding approximation, which justifies the numerical considerations below.  Figure \ref{Fig1}(b) shows the modified dispersion of exciton-polaritons with gaps both in the upper and lower polariton branch (LPB), opened by $U_{ex}$. In the following, we will concentrate on the LPB, where the condensation usually takes place. The gap and the first band where oscillations are expected have width of $\Delta_g=0.75$ $meV$ and $\Delta_1=1$ $meV$, respectively.

\begin{figure}
  \includegraphics[width=0.5\textwidth,clip]{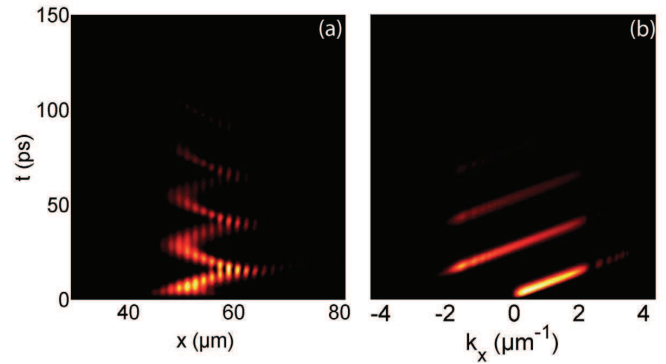}\\
  \caption{(Color online) Propagation of a photonic Gaussian pulse in the biased square-wave potential. (a) The density probability versus time and space reveals clear BOs with a period of $25.3$ $ps$ and an amplitude of $11.7$ $\mu m$, and  b) density vs time and momentum shows the characteristic reflections on the edges of the FBZ where the particles are localized. In both plots a LZT is visible as a fraction of the non-oscillating particles that escape at each period.}
\label{Fig2}
\end{figure}

To begin our analysis and demonstrate the possibility of obtaining regular single-particle BOs in such a system, we show in Fig.\ref{Fig2}(a) the propagation of a $2$-$ps$-long and $20$-$\mu m$-large photonic Gaussian pulse tuned close to the energy of the LPB at $\mathbf{k}=\mathbf{0}$, with an amplitude low enough to assume a linear regime. Under the action of the constant force and for the parameters we use, our system exhibits BOs of amplitude $A_{bo}\thickapprox12$ $\mu m$ and period $T_{bo}\thickapprox25$ $ps$; the latter is close to lifetime of the particles.  A further increase in $F$, would enhance the Landau-Zener tunneling (LZT) probability [visible in Fig.\ref{Fig2}(a)] given by ${P_{LZT}} \simeq \exp \left( { - {A^2}m^*\pi /2F} \right)$, and induce a significant splitting of the wave function at each period of oscillations. Figure \ref{Fig2}(b) shows the time evolution of the wave vector $k_x$ along the wire with characteristic reflections on the FBZ's edges, which has an extension $Z_B=2\pi/d=4$ $\mu m^{-1}$, changing $k_x$ into $-k_x$. We emphasize that the long polariton lifetime obtained in modern structures is crucial for the observation of polariton BOs. With the realistic limitations on the ramp potential, it is difficult to reduce the period of oscillations below $20$ $ps$. Therefore, the polariton lifetime should exceed this value, for the phenomenon to be observable.

\section{Bloch oscillations of the polariton BEC}
\begin{figure}
  \includegraphics[width=0.48\textwidth,clip]{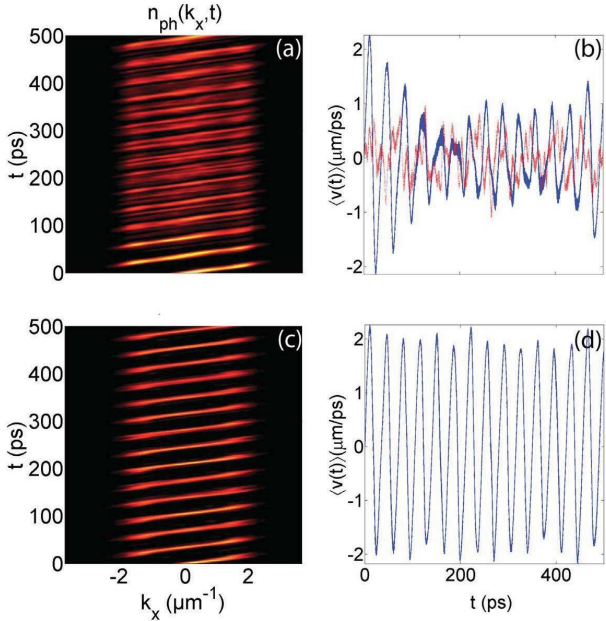}\\
  \caption{(Color online) BOs in the presence of structural disorder. (left panel) Probability density of the photonic wave function versus time and momentum. (right panel) Average velocity of the condensate versus time. (a) and (b) Non-interacting damped case ($\mu=0$) for $U_d=0.1$ $meV$ (solid blue curve) and $U_d=0.2$ $meV$ (dotted red curve). (c) and (d) Interaction-induced revival of the oscillations for $\mu=0.3$ $meV$.}
  \label{Fig3}
\end{figure}

\begin{figure}
  \includegraphics[width=0.48\textwidth,clip]{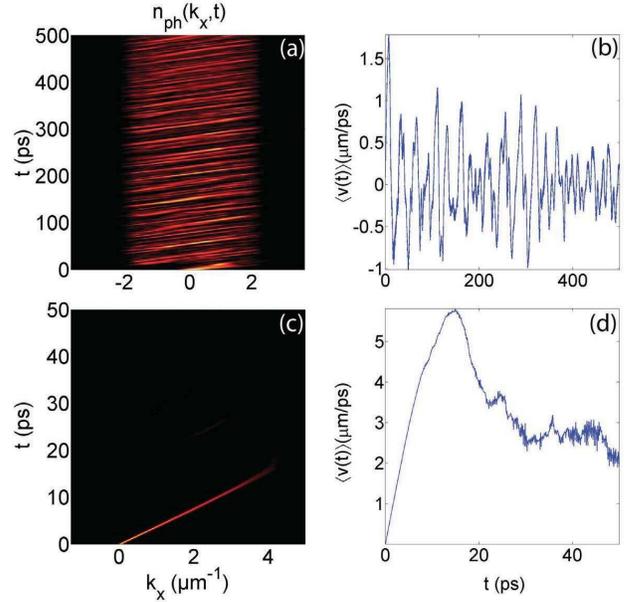}\\
  \caption{(Color online) (a)-(b) Unstable and (c)-(d) strong LZT regimes. For $\mu=0.5$ $meV$ and $3$ $meV$, respectively}
  \label{Fig4}.
\end{figure}

In the second part of this work, we consider the real specificity of the polariton condensate with respect to the coherent photonic wave, namely, the role played by interactions. This role is actually better revealed by considering the impact of a realistic structural disorder on polariton BOs. This disorder can have several origins, which could be the intrinsic sample imperfections, the non-ideal lateral etching, or some natural (artificial) fluctuations of the wells' depths and widths in the periodic potential. We assume that initially, before the potential ramp is applied, the resonantly created polariton condensate is at thermodynamic equilibrium at $T=0$ $K$, and is therefore in its lowest energy state. The ground state is found by minimizing at fixed density the free energy of the coupled exciton-photon system \cite{Malpuech2007}:
\begin{equation}\label{Energy}
    E = \int\limits_0^{{L_x}} {dx} \left\{ \begin{array}{l}
\sum\limits_{j = \left\{ {ph,ex} \right\}} {\left( {\frac{{{\hbar ^2}}}{{2{m_j}}}{{\left| {\nabla {\psi _j}} \right|}^2} + {U_j}{{\left| {{\psi _j}} \right|}^2}} \right)} \\
 + \frac{{{\Omega _R}}}{2}\left( {\psi _{ex}^*{\psi _{ph}} + \psi _{ph}^*{\psi _{ex}}} \right) + \frac{\alpha }{2}{\left| {{\psi _{ex}}} \right|^4}
\end{array} \right\}
\end{equation}
Next we solve the time-dependent Eqs.(1) and (2), starting from the initial condition given by the ground state found previously, taking into account the action of the accelerating ramp potential for different chemical potentials $\mu$ (given by the energy of the ground state). To clarify, we first assume an infinite lifetime of the particles. For a particular modulated barrier height of $4$ $meV$, to reduce the LZT probability, which decreases the number of oscillating particles, we show different accessible regimes (described in the next paragraph). Figures \ref{Fig3} and \ref{Fig4} display the probability density of the photonic wave function versus time and momentum and the average velocity of the condensate as a function of time. The latter is given by
\begin{equation}\label{vmean}
    {\left\langle v \left( t \right)\right\rangle} =  \frac{{i\hbar }}{{2{m^*}N\left( t \right)}}\int\limits_0^{{L_x}} {dx\left\{ \begin{array}{l}
 {\psi}\left( {x,t} \right)\nabla \psi ^* \left( {x,t} \right) \\
  - \psi ^* \left( {x,t} \right)\nabla { \psi}\left( {x,t} \right)
 \end{array} \right\}}
\end{equation}
where $N(t)$ the total number of particles. The latter is time dependent to compensate for absorbing boundary conditions or the lifetime of the particles. Indeed, $\left\langle v(t) \right\rangle$ is relevant for imaging the global motion of the condensate, changing its sign $\left\langle v \left( t \right)\right\rangle>0$ $(<0)$ when particles change propagation direction.

The ground state of a non-interacting or, equivalently, a very dilute condensate in the presence of disorder is mainly localized in the lowest well of the sample. As a result, while the condensate is put into motion by the constant force, it is very sensitive to the disorder; the latter induces back scattering, which leads to a dephasing and thus either to a damping of the oscillations or to their total destruction if the inhomogeneity is strong. In that latter case the system as a whole is insulating. In real systems we can expect the disorder induced by the sample imperfections to be weak with respect to the total periodic potential's amplitude. In order to introduce imperfections, we modulate the square-wave potential randomly in amplitude along the wire with a standard deviation $U_d=0.1$ $meV$ from the original potential. In Figs.\ref{Fig3}(a) and \ref{Fig3}(b) we present this non-interacting case, which corresponds to a zero chemical potential ($N=1$ particle). The dephasing is visible in Fig.\ref{Fig3}(a) where the creation of new harmonics blurs the oscillations, resulting in a damping and deformation of the motion in Fig.\ref{Fig3}(b). The solid blue curve corresponds to the parameters given here while the dotted-red curve is for a disorder that is twice stronger, showing the total suppression of the oscillations.

With the increase in the chemical potential, interactions drive the system toward superfluidity in the sense we used in Ref.\onlinecite{Malpuech2007}, that is, particles are no more affected by the presence of an in-plane potential so they can freely propagate in space without being scattered, and the interaction energy is able to efficiently screen the disorder. This phenomenon is called a "dynamical screening of disorder" in Ref.\onlinecite{Schulte}. As a result, the damping is significantly reduced. Thus, owing to the interacting nature of the particles involved, BOs of exciton-polaritons should be observable despite the presence of a structural disorder. Such a situation is depicted in Fig.\ref{Fig3}(c) and \ref{Fig3}(d) for $\mu=0.3$ $meV$.

Nevertheless, interactions are not sufficient to recover perfect oscillations obtained in linear homogeneous regimes, as one can see in Fig.\ref{Fig3}(d), and they also have drawbacks: Indeed, while the condensate is accelerated up to the FBZ, at a certain point, the velocity exceeds a critical value, and the particles enter a supersonic regime where they are no longer superfluid and are thus scattered on the disorder. Scattering results in a dephasing of the oscillations while approaching the FBZ's edges. Moreover, with a further increase in the density, parametric instability develops \cite{WuNiu, Kolovsky,Skryabin}. This kind of process is well known in the field of polaritons\cite{Savvidis,Skolnick} as the lower branch of their bare dispersion possesses an inflection point; thus above some density threshold, two polaritons can scatter at this point toward signal and idler states, conserving both momentum and energy. Here the same kind of phenomenon can take place, independently of the disorder, within the Bloch bands, which also possess inflection points. We show in Fig.\ref{Fig4}(a) and \ref{Fig4}(b) this kind of phenomenon appearing for $\mu=0.5$ $meV$; oscillations are completely deformed and no more quantifiable. It is important to note that for stronger disorders, it is not possible at all to find a screened regime [Fig.\ref{Fig3}(a)-(b)], because the system enters the parametric instability region before recovering oscillations.
\begin{figure}
  \includegraphics[width=0.5\textwidth,clip]{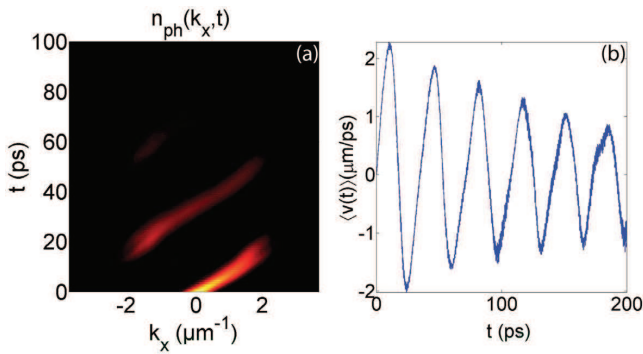}\\
  \caption{(Color online) Situation taking into account the finite lifetime of the particles for $\mu=0.35$ $meV$. (a) Oscillations remain visible, but (b) they are damped due to the decreasing interaction energy.}
  \label{Fig5}
\end{figure}

Finally, an even stronger density wipes out all the effects mentioned above and puts the Landau-Zener tunneling into play. The interactions renormalize the lowest Bloch band, closing the gap in the dispersion and increasing drastically the probability of the tunneling to the second band \cite{Lasinio}. As a result, BOs vanish, and particles are almost uniformly accelerated. They practically no longer feel the lattice. This regime is illustrated in Figs.\ref{Fig4}(c) and \ref{Fig4}(d) for a very strong, although probably not experimentally accessible, chemical potential of $\mu=3$ $meV$. The condensate is accelerated to the edge of the sample and disappears on the absorbing boundaries, as seen in \ref{Fig4}(c). The particles are no longer reflected at the FBZ edges. In \ref{Fig4}(d), the mean velocity keeps increasing until particles reach the boundary, and then it decreases because of the remaining trapped particles.

To conclude, we will now take into account the finite lifetime of the particles to move to a realistic situation. We assume that the condensate is created by a quasi-resonant pumping close to the LPB at $k_x=0$ that massively populates the ground state of our system. The chemical potential is defined by a balance between gains and losses after a stabilization time. Then we suppose that the condensate is loaded by some means in the disordered tilted periodic structure while the pump is turned off. The condensate will then oscillate a few times and exponentially die out. So we, of course, expect the interaction energy to vanish with the particles; thus, a condensate prepared with a sufficient chemical potential to screen the disorder will suffer from growing localization with decreasing density, which will damp the oscillations. So here the ratio between lifetime and oscillation period is even more crucial. Figure.\ref{Fig5} illustrates such a situation, where $\mu=0.35$ $meV$ is taken slightly higher than in Figs.\ref{Fig3}(c) and \ref{Fig3}(d) to compensate as much as possible for the localization for first oscillation periods. While the damping remains significant as seen in \ref{Fig5}(b), several oscillations should be observable as shown in \ref{Fig5}(a).

\section{conclusions}
In conclusion, we have proposed a realistic structure in which BOs of exciton-polaritons, a phenomenon not previously reported, could be observed. We have analyzed both linear and non-linear behavior and found a regime where oscillations are able to overcome the disorder expected in realistic systems as a result of to the interacting nature of the particles. At higher chemical potential, dynamical instability switches on and destroys oscillations, and at even higher density the massive Landau-Zener tunneling leads to a strong delocalization of the particles.

Regarding a real experimental investigation, we would like to point out some phenomena to be envisaged. First of all, as we mentioned in Sec.I, metallic deposition will tend to reduce the cavity photon's lifetime; thus a lateral (along the $x$-axis) square-wave etching could be considered as a serious option. In a real two-dimensional system one should expect transverse excitations to modify the instability threshold \cite{transverse}. Finally, in a configuration where Landau-Zener tunneling is significant (for a weak lattice or strong potential ramp) the wave function will split and the non-oscillating part will be backscattered at the wire's edges, which will slightly blur the motion. Nevertheless, the exceptional progress in growth and technology should lead very soon to even higher-quality samples, in which the cavity photon lifetime could approach $100$ $ps$. In such samples, Bloch-oscillations of exciton-polaritons could be observed, perhaps even in the linear regime if the structural disorder is sufficiently low.

\section{acknowledgments}
The authors acknowledge the support of the joint CNRS-RFBR PICS project and of the FP7 ITN Spin-Optronics (Grant no. 237252).


\begin{thebibliography}{99}

\bibitem{Kasprzak} J. Kasprzak et al, \emph{Nature} (London) \textbf{443}, 409-414.

\bibitem{Amo} A. Amo, J. Lefr\`{e}re, S. Pigeon, C. Adrados, C. Ciuti, I. Carusotto, R. Houdr\'{e}, E. Giacobino  and  A. Bramati, \emph{Nat. Phys.} \textbf{5}, 805 - 810 (2009).

\bibitem{Rubo} Yu. G. Rubo, \emph{Phys. Rev. Lett.} \textbf{99}, 106401 (2007).

\bibitem{LagoudakisV} K. G. Lagoudakis, T. Ostatnick\'{y}, A. V. Kavokin, Y. G. Rubo, R. Andr\'{e}, B. Deveaud-Pl\'{e}dran, \emph{Science} \textbf{326}, 974 (2009).

\bibitem{Sarchi} D. Sarchi, I. Carusotto, M. Wouters, and V. Savona, \emph{Phys. Rev. B} \textbf{77}, 125324 (2008).

\bibitem{LagoudakisJO} K. G. Lagoudakis, B. Pietka, M. Wouters, R. Andr\'{e}, B. Deveaud-Pl\'{e}dran, \emph{Phys. Rev. Lett.} \textbf{105}, 120403 (2008).

\bibitem{Savvidis} P. G. Savvidis, J. J. Baumberg, R. M. Stevenson, M. S. Skolnick, D. M. Whittaker, and J. S. Roberts, \emph{Phys. Rev. Lett} \textbf{84}, 1547 (2000).

\bibitem{Skolnick} A. I. Tartakovskii, D. N. Krizhanovskii, D. A. Kurysh, V. D. Kulakovskii, M. S. Skolnick, and J. S. Roberts, \emph{Phys. Rev. B} \textbf{65}, 081308(R) (2002).

\bibitem{Baas} A. Baas, J. Ph. Karr, H. Eleuch, and E. Giacobino, \emph{Phys. Rev. A} \textbf{69}, 023809 (2004).

\bibitem{Review} I. A. Shelykh, A. V. Kavokin, Yuri G. Rubo, T. C. H. Liew, and G Malpuech, \emph{Semicond. Sci. Technol.} \textbf{25}, 013001 (2010).

\bibitem{LagoudakisHV} K. G. Lagoudakis, T. Ostatnick\'{y}, A. V. Kavokin, Y. G. Rubo, R. Andr\'{e} and B. Deveaud-Pl\'{e}dran, \emph{Science}, \textbf{326}, 974 (2009).

\bibitem{Gippius} N. A. Gippius, I. A. Shelykh, D. D. Solnyshkov, S. S. Gavrilov, Yuri G. Rubo, A. V. Kavokin, S. G. Tikhodeev, and G. Malpuech, \emph{Phys. Rev. Lett.} \textbf{98}, 236401 (2007).

\bibitem{Wertz} E. Wertz, L. Ferrier, D. Solnyshkov, R. Johne, D. Sanvitto, A. Lemaître, I. Sagnes, R. Grousson, A. V. Kavokin, P. Senellart, G. Malpuech, and J. Bloch, \emph{Nat. Phys.} \textbf{6}, 860–864 (2010).
\bibitem{Bloch} F. Bloch, \emph{Z. Phys.} \textbf{52}, 555 (1929).

\bibitem{Wannier}  G. H. Wannier, \emph{Phys. Rev.} \textbf{100}, 1227 (1955); \emph{Phys. Rev.} \textbf{101}, 1835 (1956); \emph{Phys. Rev.} \textbf{117}, 432 (1960); \emph{Rev. Mod. Phys.} \textbf{34} , 645 (1962).

\bibitem{Mendez} E.E. Mendez F. Agull\'{o}-Rueda, and J. M. Hong, Phys. Rev. Lett 60, 2426 (1988).

\bibitem{LeoWaschkeDekorsky} K. Leo, P. Haring Bolivar, F. Br\"{u}ggemann and R. Schwedler, \emph{Solid State Commun.} \textbf{84}, 10 (1992); C. Waschke, H.G. Roskos, R. Schwelder, K. Leo, H. Kurz, and K. Kohler, \emph{Phys. Rev. Lett.} \textbf{70}, 3319 (1993); T. Dekorsky, R. Ott, H. Kurz, and K. Kohler, \emph{Phys. Rev. B}, \textbf{51} R17275 (1995).

\bibitem{Malpuech} G. Malpuech, A. Kavokin, G. Panzarini, and A. Di Carlo, \emph{Phys. Rev. B} \textbf{63}, 035108 (2001).

\bibitem{Agarwal} V. Agarwal, J. A. del R\`{\i}o, G. Malpuech, M. Zamfirescu, A. Kavokin, D. Coquillat, D. Scalbert, M. Vladimirova, and B. Gil, \emph{Phys. Rev. Lett.} \textbf{92}, 097401 (2004).

\bibitem{Dahan} M. Ben Dahan, E. Peik, J. Reichel, Y. Castin, and C. Salomon, \emph{Phys. Rev. Lett.} \textbf{76}, 4508 (1996).

\bibitem{Choi} D. I. Choi and, Q. Niu, \emph{Phys. Rev. Lett.} \textbf{82}, 2022 (1999).

\bibitem{Morsh} O. Morsch, J. H. Müller, M. Cristiani, D. Ciampini, and E. Arimondo, \emph{Phys Rev. Lett.} \textbf{87}, 140402 (2001).

\bibitem{Cristiani} M. Cristiani, O. Morsch, J. H. M\"{u}ller, D. Ciampini and, E. Arimondo, \emph{Phys. Rev. A} \textbf{65}, 063612 (2002).

\bibitem{Menotti} C. Menotti, A. Smerzi, and A. Trombenotti, \emph{New. J. Phys.} \textbf{5}, 112 (2003).

\bibitem{Deissler} B. Deissler, M. Zaccanti, G. Roati, C. D Errico, M. Fattori, M. Modugno, G. Modugno and, M. Inguscio \emph{Nature Physics} \textbf{6}, 354 - 358 (2010).

\bibitem{Ferarri} G. Ferrari, N. Poli, F. Sorrentino, and G. M. Tino, \emph{Phys Rev. Lett.} \textbf{97}, 060402 (2006).

\bibitem{Buchleinter} A. Buchleinterand, and A. Kolovsky, \emph{Phys Rev. Lett.} \textbf{91}, 253002 (2003).

\bibitem{WuNiu} B. Wu and, Q. Niu, \emph{New. J. Phys.}, \textbf{5}, 104 (2003).

\bibitem{Kolovsky} A. R. Kolovsky, H. J. Korsch, and E.-M. Graefe, \emph{Phys Rev. A} \textbf{80}, 023617 (2009).

\bibitem{AndersonSchulte} T. Schulte, S. Drenkelforth, J. Kruse, W. Ertmer, J. Arlt, K. Sacha, J. Zakrzewski, and M. Lewenstein, \emph{Phys. Rev. Lett.} \textbf{95}, 170411 (2005).

\bibitem{Modugno} G. Roati, C. D Errico, L. Fallani, M. Fattori, C. Fort, M. Zaccanti, G. Modugno, M. Modugno, and  M. Inguscio, \emph{Nature} (London) \textbf{453}, 895-898 (2008).

\bibitem{Astrakharchik} G.E. Astrakharchik, J. Boronat, J. Casulleras, and Giorgini, \emph{Phys. Rev. A}, \textbf{66}, 023603 (2002).

\bibitem{Leboeuf} M. Albert, T. Paul, N. Pavloff, and P. Leboeuf, \emph{Phys. Rev. A} \textbf{82}, 011602(R) (2010).

\bibitem{Roth} R. Roth, and K. Burnett, \emph{J. Opt. B: Quantum Semiclass. Opt.} \textbf{5}, S50 (2003).

\bibitem{Damski} B. Damski, J. Zakrzewski, L. Santos, P. Zoller, and M. Lewenstein, \emph{Phys. Rev. Lett.} \textbf{91}, 080403 (2003).

%\bibitem{Gavish} U. Gavish, and Y. Castin, \emph{Phys. Rev. Lett.} \textbf{95}, 020401 (2005).

\bibitem{Schulte} T. Schulte, S. Drenkelforth, G. K. B\"{u}ning, W. Ertmer, J. Arlt, M. Lewenstein, and L. Santos, \emph{Phys. Rev. A} \textbf{77}, 023610 (2008).

\bibitem{Drenkelforth} S. Drenkelforth, G. K. B\"{u}ning, J. Will, T. Schulte, N. Murray, W. Ertmer, L. Santos, and J. J. Arlt, \emph{New J. Phys.} \textbf{10} (2008).

\bibitem{Cho} K. Cho, K. Okumoto, N.I. Nikolaev, and A.L. Ivanov, \emph{Phys. Rev. Lett.} \textbf{94}, 226406 (2005).

\bibitem{YamamotoNature} C. W. Lai, N. Y. Kim, S. Utsunomiya, G. Roumpos, H. Deng, M. D. Fraser, T. Byrnes, P. Recher, N. Kumada, T. Fujisawa, Y. Yamamoto, \emph{Nature} (London) \textbf{450}, 529 (2007).

\bibitem{Yamamoto} T. Byrnes, P. Recher, Y. Yamamoto, \emph{Phys. Rev. B} \textbf{81}, 205312 (2010).

\bibitem{Skolnick2010} E. A. Cerda-Mendez, D. N. Krizhanovskii, M. Wouters, R. Bradley, K. Biermann, K. Guda, R. Hey, P. V. Santos, D. Sarkar, and M. S. Skolnick, Phys. Rev. Lett. \textbf{105} 116402 (2010).

\bibitem{Skryabin} A. V. Gorbach, and D. V. Skryabin, \emph{Phys. Rev. B} \textbf{82}, 125313 (2010).

\bibitem{Malpuech2007} G. Malpuech, D. D. Solnyshkov, H. Ouerdane, M. M. Glazov, and I. Shelykh, \emph{Phys. Rev. Lett.} \textbf{98}, 206402 (2007).

\bibitem{Sermage2002} B. Sermage, G. Malpuech, A.V. Kavokin, and V. Thierry-Mieg, \emph{Phys. Rev. B} \textbf{64}, 081303(R) (2001).

\bibitem{Lima} M.M. de Lima, R. Hey, P.V. Santos, and A. Cantarero, \emph{Phys. Rev. Lett} \textbf{94}, 126805 (2005).

\bibitem{Kaliteevski} M. Kaliteevski, I. Iorsh, S. Brand, R. A. Abram, J.M. Chamberlain, A.V. Kavokin and, I.A. Shelykh, \emph{Phys. Rev. B} \textbf{76}, 165415 (2007).

\bibitem{Shelykh2006} I. A. Shelykh, Y. G. Rubo, G. Malpuech, D. D. Solnyshkov, and A. Kavokin, \emph{Phys. Rev. Lett.} \textbf{97}, 066402 (2006).

\bibitem{Ciuti1998} C. Ciuti, V. Savona, C. Piermarocchi, A. Quattropani, and P. Schwendimann, \emph{Phys. Rev. B} \textbf{58}, 7926 (1998).

\bibitem{Lasinio} M. Jona-Lasinio, O. Morsch, M. Cristiani, N. Malossi, J. H. M\"{u}ller, E. Courtade, M. Anderlini, and E. Arimondo, \emph{Phys. Rev. Lett.} \textbf{91}, 230406 (2003).

\bibitem{transverse} M. Modugno, C. Tozzo, and F. Dalfovo, \emph{Phys. Rev. A} \textbf{70}, 043625 (2004).

\end{thebibliography}
\end{document}